# Blockchain Governance: An Overview and Prediction of Optimal Strategies using Nash Equilibrium

Nida Khan[1[0000-0003-3096-150X]], Tabrez Ahmad[2[0000-0001-8370-6727]], Anass Patel[3], and Radu State[1[0000-0002-4751-9577]]

[1] SEDAN Research Group, University of Luxembourg, Luxembourg
[2] ArcelorMittal Europe, Luxembourg
[3] 570Easi, France
`nida.khan@uni.lu`

**Abstract.** Blockchain governance is a subject of ongoing research and an interdisciplinary view of blockchain governance is vital to aid in further research for establishing a formal governance framework for this nascent technology. In this paper, the position of blockchain governance within the hierarchy of Institutional governance is discussed. Blockchain governance is analyzed from the perspective of IT governance using Nash equilibrium to predict the outcome of different governance decisions. A payoff matrix for blockchain governance is created and simulation of different strategy profiles is accomplished for computation of all Nash equilibria. The paper elaborates upon payoff matrices for different kinds of blockchain governance, which are used in the proposition of novel mathematical formulae usable to predict the best governance strategy that minimizes the occurrence of a hard fork as well as predicts the behavior of the majority during protocol updates. The paper also includes validation of the proposed formulae using real Ethereum data.

## 1 Introduction

Statistics forecast that the revenue generation from the blockchain industry would be equivalent to over $23.3 billion by 2023 from $2.2 billion in 2019 [1]. The expected massive growth, the envisaged disruptive nature of the technology and the inception of an entirely novel economic dimension by way of cryptoeconomics [2] has managed to garner both research and academic interest. The major proponent of blockchain inception is that it absolves the need to trust intermediaries to enable online transactions while providing auditable transparency. Blockchain technology finds diverse applications in the payments sector, asset management, supply chain management, healthcare, digital identity and Internet of Things amongst others. It has been heralded as one of the most disruptive innovations of the fourth industrial revolution [3].

Blockchain is a technological innovation that works on the concept of distributed networks requiring the coordinated efforts of multiple agents that comprise the network.

The different agents, like users, founders of the blockchain network and others have different incentives that need to be aligned to ensure a uniform and consistent progress of the blockchain network. This necessitates the requirement of a governance strategy, which consolidates the goals of the stakeholders of the blockchain network to develop incentives and strategies for the other agents in the network for providing network consistency and availability. It is predicted that an absence of an apt governance mechanism can stall and result in a sub-optimized blockchain network [4], leading to frequent soft and hard forks in the network. A hard fork, though undesirable, is a relatively cost-effective solution to disagreements, when compared to a split on a traditional structure like the government.

Blockchain governance is a subject of ongoing research and as of now no formal definition of blockchain governance exists. The paper attempts to give a formal structure to blockchain governance consolidating its position within the hierarchy of Institutional governance. The paper elaborates on the various agents involved in a blockchain platform and gives an overview of their position in the broader vision of blockchain governance. The paper discusses the three dimensions of IT governance as they apply to blockchain governance. Nash equilibrium is a concept within game theory, which has multiple applications in blockchain [5]. In this paper, Nash equilibrium is used to predict the outcome of different governance implementation strategies to find an optimal strategy that minimizes the occurrence of hard forks. The scenarios of no blockchain governance, off-chain and on-chain blockchain governance are discussed, to conclude on the best scenario that minimizes the occurrence of hard forks, the occurrence of which is unavoidable in many cases.

The paper is a pioneer in giving a multidisciplinary view of blockchain governance and using game theory concepts to analyze different governance strategies. Related work is given in section 2. A multi-faceted view of blockchain governance is given in section 3 while section 4 gives the relevant background on Nash equilibrium. Application of Nash equilibrium and simulations of the developed blockchain governance strategic game are accomplished to conclude on optimal strategy profiles in section 5. Pay-off matrices are constructed for off-chain and on-chain blockchain governance and mathematically analyzed to derive formulae that can help in prediction of hard forks based on the choice of governance decision during a protocol update in subsections 5.3 and 5.4. Data from Ethereum fork is analyzed to verify the developed formulae in section 6. The conclusion of the undertaken study is given in section 7.

## 2   Related Work

Davidson *et al.* view blockchain governance from the perspective of Institutional governance [6]. Beck *et al.* focus on a research framework agenda for blockchain governance, deriving rules from IT governance [7]. Chohan discuss the governance issues in decentralized autonomous organizations [8]. Atzori elaborates on the key points of blockchain governance from a political perspective, focusing on the distinguishing characteristics from State authority, citizenship and democracy [9]. Reijers *et al.* do a comparative analysis of blockchain governance, espoused by blockchain developers with the governance concepts discussed within social contract theories [10]. Barrera and Hurder view blockchain policy upgrade as a coordination game and develop a simple model of strategic chain choice [11]. Arrunada and Garicano discuss that new forms of soft governance need to be developed that allow the decentralized network to avoid bad equilibria [12]. In this paper, we discuss the hierarchy of Institutional governance within which blockchain governance resides and analyze IT governance as it applies in a blockchain organization. The implementation categories of blockchain governance is highlighted and Nash equilibrium is used to construct payoff matrices to predict the best implementation strategy for blockchain governance.

## 3      Blockchain Governance

Blockchain is a distributed, decentralized network that came into inception with the launch of Bitcoin. The coordination of various entities that comprise a blockchain network to ensure that the network functions without conflicts to accomplish the strategic goals of the network requires a **governance** mechanism. The main components that comprise the blockchain network can be summarized as follows:

1. Validators
2. Users
3. Consensus Protocol
4. Governance Mechanism

A blockchain network is analogous to an organization like Google or Microsoft, with its own set of users, employees (validators), protocols and a governance structure. Consider the email services, Gmail provided by Google and Outlook by Microsoft, where their purpose is the same but the governance mechanisms of the two corporations, Google and Microsoft, differ. Services provided by a blockchain network, like cryptocurrency payments or dapps, can be comparable to the services provided by a technology company. The blockchain network built by an organization, example Stellar, will put it in the category of a technology company. Presently there are numerous organizations like Bitcoin, Ethereum and Tezos amongst others, providing a blockchain-based infrastructure. The organizations differ in their consensus protocols, governance mechanisms, services and strategic goals. Thus, a universal governance mechanism cannot exist for blockchain in the wake of multiple service providers for the technology.

The blockchain network facilitated by an organization will make the respective organizations be analogous to a technology company, where the standards and frameworks for corporate governance would be applicable. This view is corroborated by the domain of corporate governance, which indicates that *"Corporate governance includes all types of firms whether or not they are formed under civil or common law, owned by the government, institutions or individuals, privately or publicly traded."* [13]. This implies that firms like Stellar will be in the realm of corporate governance. Corporate governance is concerned with the relationship among the many players in an organization and the organizational goals for which governance is needed. The different players in an organization are the stakeholders, management, board of directors, employees, suppliers, customers, lenders, regulators, the environment and the community at large [14]. Corporate governance deals with the set of rules, processes and practices which aid in giving the organization a direction and a framework to ensure its functioning is within the defined parameters. Corporate governance lies within the realm of different Institutional governance systems [15]. Blockchain governance, ensuing from corporate governance would essentially follow the similar set of processes and rules to achieve the vision of the founders while ensuring that the participants in the network like the validators and users contribute to its goals. A good governance protocol would ensure that the blockchain would be environmentally friendly, would minimize the occurrence of forks and changes in the code would represent the agreement of the majority of the agents in the blockchain network.

### 3.1 IT Governance

Information Technology (IT) governance is a critical driver for corporate governance. It is a component of corporate governance but the relationship between the two remains largely unexplored, despite the acknowledgement that an organization would fail to accomplish its goals without a good IT governance mechanism [16]. Weill and Ross describe IT governance as *"specifying the decision rights and accountability to encourage desirable behavior in the use of IT"* [17]. This definition includes establishment of a set of processes and deciding the authorities for providing the input to making decisions. They indicate a set of questions that must be addressed for an effective IT governance mechanism, which includes the decisions that must be made for effective management and use of IT, delegation of authority for decision-making and finally the process of decision-making coupled with monitoring. Simonsson and Johnson pointed out that a shared definition is lacking in the field of IT governance [18]. They provided an IT governance definition based on consolidation of literature, which identified three dimensions namely, the ***domain***, ***processes*** and ***scope*** in which IT decisions are made and carried out. In this paper we proceed with the three dimensions that encompass IT governance stated in [18] and justify the usage by ensuring conformity to the questions addressed by Weill and Ross in [17]. The paper focuses on IT governance as blockchain organizations are primarily technology initiatives increasing the significance of IT governance in the context of blockchain platforms.

**Domain.** The *domain* dimension includes the entities, the decisions should consider and comprise of goals, processes, people and technology [18]. In the context of blockchain technology, goals of any public blockchain technology majorly encompass the characteristics of a blockchain platform, which are as follows:

1. Trustless: The technology does not depend on intermediaries and third parties for conducting transactions.
2. Immutable: The ledger is incorruptible, being verified by innumerable nodes and cryptography ensures a permanent record of transactions.

3. Decentralization: The technology is not controlled by a single entity and power is distributed.
4. Availability: The technology does not have a single point of failure.
5. Auditability: The immutable blockchain ledger is verifiable by anybody.
6. Confidentiality: The technology is public, yet the identities of the users remain pseudonymous.

The goals outlined above might differ in some parameters in case of consortium and private blockchain networks. Additional goals of development of advanced protocols for enhancement of performance, monitoring of protocol updates or changes and alignment to the goals of the stakeholders also exist.

Examples of processes in blockchain include the implementation and management of peer to peer network, the distributed ledger, cryptographic algorithms involved, code development and prompt error corrections on detection. The people involved are the core developers that create the code to manage the functioning of blockchain technology, full nodes that maintain a copy of the entire blockchain ledger to put the code into action, the organization, which manages the funds and reimburses the core developers and the users, who use the blockchain network. The organization can be for profit like Ripple, or non-profit like Stellar. Technology broadly includes the hardware required and the software that aids in the setup and implementation of the blockchain network.

**Decision-making process.** The second dimension of IT governance is the *decision-making process* [18]. Decision-making process involves three phases where the matter under consideration is thought over, deliberated and then transformed into a model. The model is investigated and then implemented in the organization, which is monitored for performance evaluation. In the context of blockchain governance, the first phase can be considered equivalent to the protocol submission in Ethereum as Ethereum Improvement Proposal (off-chain governance (subsection 3.2)) or proposal submission in Tezos (on-chain governance (subsection 3.2)). The protocol can then be implemented in the testnet as in Tezos and voting ensues to deliberate upon its acceptance. If a majority is reached in the voting process involving the decision-makers, the protocol is implemented in the main blockchain network and monitored for any software bugs. Thus, the process is similar to IT governance and a formal framework would necessitate the adherence to all the three phases in order.

**Scope.** The last and third dimension in the definition of IT governance involves a discussion on the *scope*. Every governance decision implies a long and a short-term aspect with a correlation between the timeline of the decision and the level at which it is made [18]. The scope dimension encompasses tactic and strategic decisions. In the context of a blockchain network, tactic governance decisions can be changing the user interface of the blockchain network whereas a strategic decision would involve a protocol update.

A brief discussion of the three dimensions of an IT governance framework as it applies to designing an effective blockchain governance mechanism indicates that the questions that need to be addressed for an effective governance framework as per Weill and Ross [17] have been majorly targeted. However as per the definition of IT governance *"to encourage desirable behavior in the use of IT"*, there is a clear indicator on the use of incentives, which can be both monetary and non-monetary. The emphasis on incentives is less in this definition but they have been recognized as pivotal in the design and evaluation of any information system [19]. When the blockchain network is designed such that the users are encouraged and incentivized to use the network as per the design objectives then a very important design goal is accomplished. This incentive alignment can be derived from the theory of Nash equilibrium [20] or the principles of game theory as they apply to institutional economics [21]. **Fig. 1** represents the hierarchy of governance in blockchain.

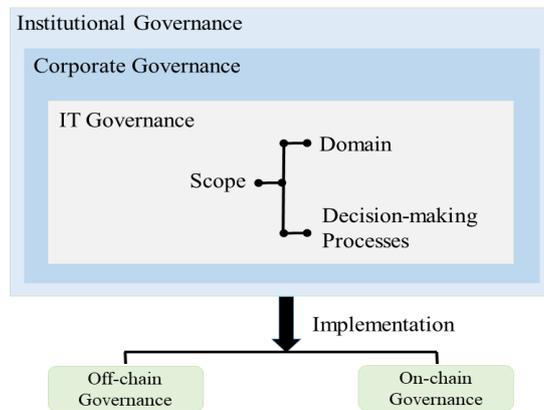

**Fig. 1.** Hierarchy of Governance in Blockchain

### 3.2 Implementation Methodologies for Blockchain Governance

The implementation of blockchain governance can be either off-chain or on-chain. Off-chain governance is seen in Bitcoin and Ethereum, which is characterized by an informal decision-making process independent of the underlying blockchain code base. Off-chain governance resembles the traditional governance mechanisms with relative centralization and distribution of power between the blockchain agents. Users who lack the technical expertise or the requisite financial resources fail to contribute to the governing decisions. The off-chain design mechanism for governance is driven by the belief that governance is an unpredictable and emergent phenomenon, which cannot be hardcoded into the blockchain in advance. The protocol updates are submitted by the core developers in the form of formal improvement proposals, for example as Bitcoin Improvement Proposals (BIPs) in Bitcoin [22]. The e-proposals are then deliberated through social media and discussion groups. One of the issues with off-chain governance is lack of incentives for the proposers, which can lead to a small group of developers submitting proposals and hence centralization.

On-chain governance embodies the spirit of decentralization in the governance mechanisms as seen in Tezos. The concentration of power witnessed in off-chain governance strategy is minimized and the governance is implemented by virtue of a series of processes as opposed to a simple majority consensus. The exact procedure in on-chain governance might vary between different blockchain platforms. The proposals are submitted on-chain, voted and if agreed upon are deployed on the blockchain testnet for a designated amount of time. If the final vote is in favor after the testnet deployment, then the proposal is incorporated in the main blockchain. In order to achieve the settlement, governance rules are written into code and are a part of the blockchain enhancing transparency. It also reduces the turnaround time for a proposed protocol update. The infrastructure required for a well-functioning on-chain governance is huge comparatively and research is in progress for the requisite coordination tools that facilitate an efficient communication in on-chain governance [23].

## 4   Nash Equilibrium

Non-cooperative games in game theory are those where competition exists amongst the individual players and it is characterized by an absence of rules that enforce cooperative behavior. Blockchain platforms can be viewed as non-cooperative games between the validators, who compete to add the next block to gain an economic advantage. Nash equilibrium can be used to derive a solution for non-cooperative games and can be utilized to aid in predicting optimum equilibria in blockchain governance [11]. In the mathematical context, if it is proved that a Nash equilibrium exists then this is equivalent to proving that a solution exists for a fixed-term problem [24]. Nash equilibrium introduced the concept of games with *n* participants, *players*, where each would need to decide on a course of action, *strategy*. Strategies can be *pure* or *mixed*. A pure strategy gives the moves a player will make during the course of the game under any situation. In blockchain governance, it represents the set of all possible choices a player can make. Mixed strategies are probability distributions over decisions as in the context of choosing a proposal, different agents in the blockchain network will choose one from the available choices ensuring that all the choices are taken by a random proportion of the agents. A pure strategy is analogous to a degenerate case of a mixed strategy, where the concerned pure strategy is chosen with a probability **1**, whereas all others are chosen with a probability of **0**.

Nash equilibrium can be stated as a *"set of strategies, one for each of the n players of a game, that has the property that each player's choice is his best response to the choices of the n-1 other players"* [20] [25]. A payoff matrix is used to represent the available strategies and the players, where each cell of the matrix gives the outcome of different choices by each player. The matrix gives the outcome of an individual's choice of strategy in terms of gain or loss that a player undergoes, when his choice of strategy is executed, given the choice of other players. Thereafter, the matrix is analyzed to determine optimal strategies. John Nash in [25] defined that a *n-person game* is a set of *n players* or *positions*, where each player, *i*, has a finite set of pure strategies and a payoff function, $p_i$ mapping the entire set of *n-tuples* of pure strategies into real numbers. The term *n-tuples* implies a set of *n* items, where each item is associated with a different player [25]. The mixed strategy of a player is a collection of positive numbers that are in one to one correspondence with his pure strategies and have unit sum. The payoff function $p_i$ is characterized by a unique extension to the *n*-tuples of mixed strategies and is linear in the mixed strategy of each player. The equilibrium point is a *n*-tuple where each player's mixed strategy maximizes his payoff if the strategies of other players is not changed. Thus, at equilibrium point each player's strategy is optimal against those of others [25].

### 4.1 Pareto Optimality

In game theory if a solution is Pareto optimal, then this implies that the strategy profile is such that there can be no other strategy in which a player's payoff can be increased, without decreasing the payoff of at least one other player. A strong Nash equilibrium, which implies stability against unilateral deviations of players and also against unilateral deviations of any subcoalition of players, is Pareto optimal [26].

## 5 Application of Nash Equilibrium to Blockchain Governance

Blockchain governance has some distinctive features differentiating it from traditional corporate governance processes, as given below:

- Decentralized nature of governance, where the main role of the governance mechanism is to steer the community to achieve certain outcomes without having explicit levers to achieve these outcomes.
- Possibility of forks or opting out if the outcome is not aligned with individual goals is relatively hard wired in the whole governance process unlike the corporate setting where the cost benefit of opting out are quite high and in many cases might imply starting completely afresh.
- The decision of the individual on a blockchain platform is based not just on an individual choice but the impact that it is likely to have on the entire blockchain network, through the community's collective decision, resulting in a network effect.

We have applied Nash equilibrium to blockchain governance and simplified the governance process as a two-player strategy game, where voters **V** are a group of **k** entities involved in the voting of acceptance and rejection of a proposal, which is the decision-making process when viewed from the perspective of IT governance. The impact of the decision will be on the broader community **C** connected to the blockchain and comprising of **n** entities. The payoff matrix for our evaluation is given in **Fig. 2**. **β** represents the proportion of voters, **V**, who voted to accept the proposal by a **Yes** whereas **γ** represents the proportion of the broader community moving to the upgraded chain once a proposal is accepted based on the outcome of the voting process. V comprises of vali-dators, who validate the transactions in blockchain, developers, blockchain token holders, blockchain stakeholders and any user of the blockchain platform who participate in the voting process of any new blockchain proposal under consideration. C consists of all blockchain users who will be affected by the decision, including V.

|  | | Community (C) | |
| --- | --- | --- | --- |
|  | | Upgraded Chain | Original Chain |
| Voters (V) | Yes | β S(V), γ S(C) | β S(V), (1-γ) S(C) |
|  | No | (1-β) S(V), γ S(C) | (1-β) S(V), (1-γ) S(C) |

**Fig. 2.** Payoff Matrix for Blockchain Governance Game

Whenever any proposal to update the blockchain code is initiated, the individual members of **V** must decide if they want to vote to accept the proposal, **Yes**, or reject the same, **No**. Once the voting cycle for a new proposal is over, and there is no uniform consensus, the individual (in larger C) must choose between two chains: stay on the original chain **O** or join the new upgraded chain with an updated policy **U**. The payoff matrix given in **Fig. 2** enumerates two strategies, *Yes* and *No* for **V** as described formerly and **C** also has two strategies, which are *Upgraded Chain* and *Original Chain*.

$$V(k) = \{Y, N\} \subseteq C(n) = \{O, U\} \tag{1}$$

Payoff (**S**) is the cumulative of individual payoffs as a result of the decision. We have assumed no distinction between entities in the Voter, (V) and Community, (C) in the value of payoffs ($S_v$, $S_c$):

$$S(V) = \sum_{i=1}^{k} S(V_i) = kS_v \tag{2}$$

$$S(C) = \sum_{i=1}^{n} S(C_i) = nS_c \tag{3}$$

### 5.1 Simulation of Blockchain Governance Game

We use the software tools for game theory, version 15.1.1, provided by the Gambit project [27] to simulate a 2 player strategy game for **V** and **C**, the payoff matrix of which is given in **Fig. 2**. Gambit was used to compute all possible Nash equilibria for different values of **β** and **γ** and the computation results are depicted in **Table 1**. Player 1 is **V** and player 2 is **C** in **Table 1**. An analysis of the computations in **Table 1** is given below:

- Simulation 1. There is only one Nash equilibrium, where 100% of V vote 'Yes' to accept the proposal and the entire broader community C moves to the upgraded chain.
- Simulation 2. There is only one Nash equilibrium where 100% of V vote 'No' to reject the proposal and the entire broader community C stays on the original chain.
- Simulation 3. There is only one Nash equilibrium where 100% of V vote 'Yes' to accept the proposal but entire C stays on the original chain, ignoring the outcome of the voting process.
- Simulation 4. There is only one Nash equilibrium where 100% of V vote 'No' to reject the proposal but entire C still move to the upgraded chain, ignoring the outcome of the voting process.
- Simulation 5. There are 4 Nash equilibria. Equilibria 1 and 2 indicate that when 50% of V vote as 'Yes' to accept the proposal there are two optimal decisions for C, where C can either stay on the original chain or move to the upgraded chain. Equilibria 3 and 4 indicate that when 50% of V vote 'No' to reject the proposal, the optimal decision for C will again be that the community stays on the original chain or moves to the upgraded chain. In the equilibria for this simulation, the community will be divided.

Table 1. Simulation of Blockchain Governance Game

| Simulation # | β, γ | Nash Equilibria # | 1: Yes | 1: No | 2: Upgraded Chain | 2: Original Chain | V Payoff | C Payoff |
|---|---|---|---|---|---|---|---|---|
| 1 | 1, 1 | 1 | 1 | 0 | 1 | 0 | 1 | 1 |
| 2 | 0, 0 | 1 | 0 | 1 | 0 | 1 | 1 | 1 |
| 3 | 1, 0 | 1 | 1 | 0 | 0 | 1 | 1 | 1 |
| 4 | 0, 1 | 1 | 0 | 1 | 1 | 0 | 1 | 1 |
| 5 | 1/2, 1/2 | 1 | 1 | 0 | 1 | 0 | 1/2 | ½ |
|  |  | 2 | 1 | 0 | 0 | 1 |  |  |
|  |  | 3 | 0 | 1 | 1 | 0 |  |  |
|  |  | 4 | 0 | 1 | 0 | 1 |  |  |
| 6 | 3/5, 7/10 | 1 | 1 | 0 | 1 | 0 | 3/5 | 7/10 |
| 7 | 1/5, 2/5 | 1 | 0 | 1 | 0 | 1 | 4/5 | 3/5 |
| 8 | 7/10, 1/5 | 1 | 1 | 0 | 0 | 1 | 7/10 | 4/5 |
| 9 | 7/20, 18/25 | 1 | 0 | 1 | 1 | 0 | 13/20 | 18/25 |

- Simulation 6. In this the value of [β, γ > 0.5]. There is only one equilibrium where 60% of V vote 'Yes' to accept the proposal and 70% of C move to the upgraded chain.
- Simulation 7. The value of [β, γ < 0.5] and there is only one Nash equilibrium. When 80% of V vote 'No' to reject the proposal, 60% of C stay on the original chain.
- Simulation 8. The value of β > 0.5 and γ < 0.5. There is only one Nash equilibrium where despite 70% of V voting 'Yes' to accept the proposal, 80% of C stay on the original chain, ignoring the outcome of the voting process.
- Simulation 9. The value of β < 0.5 and γ > 0.5. There is only one Nash equilibrium where 35% of V vote 'Yes' to accept the proposal, 65% of V vote 'No' and 72% of C move to the upgraded chain, ignoring the outcome of the voting process.

The above analyses can be concluded with the following observations:

1. The broader community, C can take a decision independent of the outcome of the voting process.
2. If the proportion of V voting 'Yes' or 'No' to a proposal is known, the payoff matrix in Fig. 2 can be used to predict the optimal outcomes for C.
3. When C is divided on the outcome as in simulation 5, then there will always be more members on either the original or the upgraded chain increasing the payoff of that chain.

### 5.2 Evaluation of Different Kinds of Blockchain Governance

We evaluate three possible scenarios of blockchain governance. We analyze the case of no governance and then investigate the performance of off-chain and on-chain governance mechanisms on the payoff to the community under specific conditions. We derive mathematical formulae based on our evaluation to aid in prediction of optimal strategies for blockchain governance and make the following assumptions for our analysis:

1. $\beta$ is the proportion of the voters **V**, voting **Yes** to accept a new proposal and can be extended to represent the entire community **C**.
2. The proposal is accepted if the majority of voters vote **Yes** in favor of the proposal. This implies that $0.5 < \beta \leq 1$.
3. If $\beta = 0.5$, then the above majority rule in assumption **2** will apply.
4. The broader community **C** moves to the upgraded chain, **U**, if voting was in majority for the proposal acceptance as indicated by assumptions **2** and **3**.
5. C moving to the upgraded chain, **U**, indicates that $0.5 \leq \gamma \leq 1$.
6. Only two possibilities exist for the entire community, which is that they either stay on the original chain, **O** or move to the upgraded chain, **U** post the decision following the voting process. We have ignored that a section of the community can create a new hard fork, separate from the original and upgraded chain.

**No Governance.** In the scenario of no governance each **C** member is choosing the best possible option for their own benefit whenever any change is made to the chain. The individual decision-making is complicated by a high degree of opacity in information exchange on proposed changes, relative payoff to the community and the possibility of hard fork occurring as a result of the change. Let $\gamma$ be the proportion of **C** which chooses to move to the upgraded chain while **(1- γ)** choose to remain on the original chain. The payoffs are reduced on both chains as compared to single-chain equilibria by a factor **[γ, (1-γ)]** on **U**, **O** chains respectively. In equation 4, $S_U$ represents total payoff in upgraded chain and in equation 5, $S_O$ represents total payoff in original chain.

$$S_U = \gamma n S_C \leq n S_C \qquad (4)$$

$$S_O = (1 - \gamma)nS_C \leq nS_C \tag{5}$$

**Off-chain Governance. Fig. 3(a)** depicts the off-chain governance mechanism (see subsection 3.2) with proportion **β** of voter community voting *yes* if there is no outright consensus. Depending on the majority preference, the chain is upgraded or the original is maintained. After the outcome is known, everyone has an option of staying on the upgraded chain or the original chain. The voting process is a group exercise while the hard fork decisions are individual. In this case with the information available, the member is more likely to go with the majority decision in order to maximize payoff, hence the probability of hard fork is lower than the previous case of no governance.

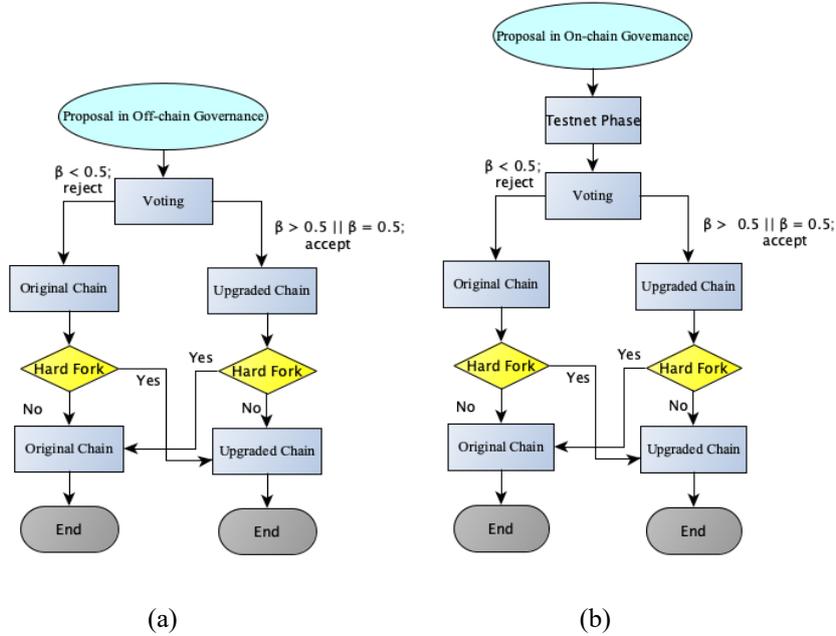

(a)    (b)

**Fig. 3.** Flowchart for Off-chain and On-chain Blockchain Governance

**On-chain Blockchain Governance. Fig. 3(b)** depicts the on-chain governance mechanism (see subsection 3.2) where the proposal is deployed on the testnet for voter community to use for a certain time period. The mechanism defers from the simple majority voting as in this case the objective is payoff maximization through maximum consensus. Thus, a vocal minority which feels strongly about any proposed change has the possibility to convince the community during the testnet phase. If there is no uniform consensus after testnet phase, the proposals are put to vote with majority chain being chosen. The possibility of hard fork will remain but with an additional exchange round the probability of **C** group member going against the majority consensus reduces as compared to both the previous cases of no governance and off-chain governance.

### 5.3   Off-chain Governance Payoff Matrices

We construct payoff matrices for off-chain blockchain governance using the basic blockchain governance matrix given in **Fig. 2**. We constructed 3 matrices of dimensions 2×2 for the respective values of β, γ indicated in the different rows of **Fig. 4**. We selected rows from the constructed matrices, which corresponded to the assumptions listed in subsection 5.2 and depicted them in a new matrix in **Fig. 4**. The matrix can be explained with the following strategy for **V, C**:

- If **V** validates the proposal as accepted then the chain is upgraded, the payoff is maximized and there is no threat of hard fork as shown in the cell, **A1** of the evaluation matrix. This is a pure strategy equilibrium. Equation 6 gives the surplus payoff for the voters S(V), while equation 7 gives the surplus payoff for the community S(C).

|  | Community (C) | |
|---|---|---|
| Voters (V) | Upgraded Chain | Original Chain |
| Accept (β, γ = 1) | **A1** S(V), S(C) | **A2** |
| Accept (0.5 < β, γ < 1) | **B1** β S(V), γ S(C) | **B2** (1-β) S(V), (1-γ) S(C) |
| Reject (0 < β, γ < 0.5) | **C1** β S(V), γ S(C) | **C2** (1-β) S(V), (1-γ) S(C) |

**Fig. 4.** Evaluation Matrix for Off-chain Blockchain Governance

$$S(V) = S_{Yes} - S_{No} = kS_V \tag{6}$$

$$S(C) = S_U - S_O = nS_C \tag{7}$$

- If **V** does not validate the proposal but after a round of voting, majority chooses to go with the proposal there is risk of hard fork with members of **C**, who may feel the decision as tyranny of majority choosing to remain on the original chain. However surplus maximization still exists with upgraded chain, as depicted in cell **B1** of the payoff matrix, as [0.5 < β, γ < 1]. This is also Pareto optimal as compared to the original chain, **B2**, as members moving to alternate chain to increase their payoff always results in decreasing the payoff for others. Equation 8 gives the surplus payoff for the voters S(V), equation 9 gives the surplus payoff for the community and

14equation 10 gives the total surplus payoff **S**. When we compare the mixed strategy equilibria for cells **B1** and **B2**, we observe that the payoff in **B1** is more than in **B2**.

$$S(V) = S_{Yes} - S_{No} = \beta k S_V - (1-\beta)k S_V = (2\beta - 1)k S_V > 0 \tag{8}$$

$$S(C) = S_U - S_O = \gamma n S_C - (1-\gamma)n S_C = (2\gamma - 1)n S_C > 0 \tag{9}$$

$$S = (2\beta - 1)k S_V + (2\gamma - 1)n S_C > 0 \tag{10}$$

- If **V** does not validate the proposal but after a round of voting, majority chooses to stay on the original proposal there is again a risk of hard fork with some members of **C** choosing to go on the updated chain. In this case surplus maximization will be with original chain, **C2** as [0.5 < 1-β, 1-γ < 1]. This is also Pareto optimal as compared to the original chain, as seen in **C1** members moving to alternate chain to increase their payoff always results in decreasing the payoff for others. Equation 11 gives the surplus payoff for the voters S(V), equation 12 gives the surplus payoff for the community S(C) and equation 13 gives the total surplus payoff. Similar to the argument in the comparison of the payoff in **B1** and **B2**, the mixed strategy equilibria in **C2** when compared with the mixed strategy equilibria in **C1** has a higher payoff.

$$S(V) = S_{No} - S_{Yes} = (1-\beta)k S_V - \beta k S_V = (1 - 2\beta)k S_V > 0 \tag{11}$$

$$S(C) = S_O - S_U = (1-\gamma)n S_C - \gamma n S_C = (1 - 2\gamma)n S_C > 0 \tag{12}$$

$$S = (1 - 2\gamma)n S_C + (1 - 2\beta)k S_V > 0 \tag{13}$$

### 5.4 On-chain Blockchain Governance Payoff Matrices

We construct payoff matrices for on-chain blockchain governance using the basic blockchain governance matrix given in **Fig. 2**. We constructed 3 matrices of dimensions 2×2 for the respective values of β, γ indicated in the different rows of **Fig. 5**. We selected rows from the constructed matrices, which corresponded to the assumptions listed in subsection 5.2 and depicted them in a new matrix in **Fig. 5**. The matrix can be explained with the following strategy for **V**, **C**:

- If **V** validates the proposal as accepted then the chain is upgraded, the payoff is maximized and there is no threat of hard fork as depicted in the cell, **A1** of the evaluation matrix. This is a pure strategy equilibrium. Equation 6 gives the surplus payoff for the voters S(V), while equation 7 gives the surplus payoff for the community.
- If **V** does not validate the proposal and even after the testnet round there is no consensus, then the voting takes place with the majority choosing to go with the proposal. The risk of hard fork is still there but is reduced as compared to off-chain as there has been a round of consultation and refinement. The surplus maximization is with upgraded chain, **B1**, as [0.5 < β, γ < 1]. This is also Pareto optimal as compared to the original chain, **B2**. Equation 8 gives the surplus payoff for the voters S(V), equation 9 gives the surplus payoff for the community and equation 10 gives the total surplus payoff S. When we compare the mixed strategy equilibria for cells **B1** and **B2**, we observe that the payoff in **B1** is more than in **B2**.

|  | Community (C) | |
|---|---|---|
| Voters (V) | Upgraded Chain | Original Chain |
| Accept ($\beta, \gamma = 1$) | A1<br>S(V), S(C) | A2 |
| Accept ($0.5 < \beta, \gamma < 1$) | B1<br>$\beta$ S(V), $\gamma$ S(C) | B2<br>$(1-\beta)$ S(V), $(1-\gamma)$ S(C) |
| Reject ($0 < \beta, \gamma < 0.5$) | C1<br>$\beta$ S(V), $\gamma'$ S(C) | C2<br>$(1-\beta)$ S(V), $(1-\gamma')$ S(C) |

**Fig. 5.** Evaluation Matrix for On-chain Blockchain Governance

- If after the voting, post testnet round, the majority chooses to remain with the original proposal, there is a possibility that the upgraded proposal will be payoff maximizing because of vocal minority with intense preference can convince more **C** members to go in for the upgraded chain $\gamma' > \gamma$, thus increasing the total community payoff. The risk of hard fork is there, based on proportion of **C** members who prefer coordinating to maximize payoff, when compared to those who wish to go with the majority opinion. As $\gamma'$ increases more members shift to upgraded chain thus neutralizing the majority vote. Equation 14 gives the surplus payoff for the voters S(V), equation 15 gives the surplus payoff for the community S(C) and equation 16 gives the total surplus payoff, S. If S > 0, majority will move to the upgraded chain as seen in equation 17. If S < 0, then the majority will stay on the original chain as seen in equation 18.

$$S(V) = S_{Yes} - S_{No} = \beta k S_V - (1-\beta)k S_V = (2\beta - 1)k S_V < 0 \qquad (14)$$

$$S(C) = S_U - S_O = \gamma' n S_C - (1-\gamma')n S_C = (2\gamma' - 1)n S_C > 0 \qquad (15)$$

$$S = (2\gamma' - 1)n S_C + (2\beta - 1)k S_V \qquad (16)$$

$$S > 0, \text{ if } (2\gamma' - 1)n S_C > -(2\beta - 1)k S_V = (1 - 2\beta)k S_V \qquad (17)$$

$$S < 0, \text{ if } (2\gamma' - 1)n S_C < (1 - 2\beta)k S_V \qquad (18)$$

## 6  Application of the Proposed Mathematical Formulae to Real Data

We use Ethereum data to verify our proposed mathematical formulae for prediction of optimal strategies. Ethereum has off-chain governance and after the DAO hack there was a proposal to upgrade the chain in 2016. Major mining pools, amounting to 54%, supported the decision by voting 'Yes' [28] making $\beta = 0.54$. This scenario corresponds to the 2$^{nd}$ row of the evaluation matrix in **Fig. 4** with cells B1 (upgraded chain) and B2 (original chain). As per our analysis in subsection 5.3, the risk of hard fork exists. However, cell B1 represents a Pareto optimal condition with surplus maximization existing with the upgraded chain, where the payoff of B1 is more than B2 as validated by equations 8, 9 and 10. This will result in the community moving to the upgraded chain (B1). As it was witnessed, a decision was taken by the Ethereum community to go for the protocol update and this resulted in a split in the community C, creating the hard fork Ethereum (upgraded chain). The present day metrics of the upgraded chain (Ethereum) [29] and the original chain (Ethereum Classic) [30] is a validation of our predictions in equations 8, 9 and 10 indicating that the payoff is more with the upgraded chain (Ethereum), where the majority of the community moved to.

## 7  Conclusion

In this paper we analyzed the position of blockchain governance in the hierarchy of Institutional governance to derive a formal structure for blockchain governance. The paper viewed blockchain governance from the dimensions of IT governance and then analyzed one dimension of IT governance, namely decision-making process as in the form of voting on a new blockchain improvement proposal, by using Nash equilibria to predict optimal governance strategies. The objectives of the governance process may vary with the blockchain platform, its community composition and business logic. A game theory simulation of the respective strategies of the players was used in an attempt to define the most optimum scenarios based on choices made by the voters or the broader community. The paper, through an analysis of the payoff matrices, provides mathematical formulae that can predict the occurrence of a hard fork on acceptance/ rejection of a new proposal as well as give an indication whether the majority of the community will move to the upgraded chain or stay on the original chain. We tried to maximize the payoff for the community while steering them to desired outcomes in our analyses and it was observed that a split in the community through a hard fork diminishes the payoff. Off-chain governance mechanism tries to avoid the potential loss of payoff through an inclusive voting process while on-chain includes a pilot testing with an inclusive voting process. On-chain provides an added opportunity to change the value of $\gamma$ and hence can be concluded as a better governance proposition.

## 8 Acknowledgments

This work is funded by the Luxembourg National Research Fund under its AFR-PPP Programme (FNR11617092), which is aimed at providing PhD and Post Doc grants for innovation and industry partnerships. The authors would like to thank Dr. Zsofia Kraussl from Luxembourg School of Finance for her guidance in the description of blockchain governance.

**References**

1. Statista, Size of the blockchain technology market worldwide from 2018 to 2023, https://www.statista.com/statistics/647231/worldwide-blockchain-technology-market-size/, last accessed 2019/9/7.
2. Dannen, C.: Cryptoeconomics Survey. In: Introducing Ethereum and Solidity, pp. 139-147. Springer, Berkeley, CA (2017).
3. Pollock, D.: The Fourth Industrial Revolution Built On Blockchain And Advanced With AI, https://www.forbes.com/sites/darrynpollock/2018/11/30/the-fourth-industrial-revolution-built-on-blockchain-and-advanced-with-ai/#7a756b94242b, last accessed 2019/12/5.
4. Filippi, P., Mcmullen, G.: Governance of blockchain systems: Governance of and by Distributed Infrastructure. (Research Report) Blockchain Research Institute and COALA (2018).
5. Liu, Z., Cong, N., Wang, W., Niyato, D., Wang, P., Liang, Y., Kim, D..: A Survey on Applications of Game Theory in Blockchain, Cornell University (2019).
6. Davidson, S., Filippi, P., Potts, J.: Disrupting Governance: The New Institutional Economics of Distributed Ledger Technology. SSRN Electronic Journal (2016).
7. Beck, R., Müller-Bloch, C., King, J.: Governance in the Blockchain Economy: A Framework and Research Agenda. Journal of the Association for Information Systems (19), 1020–1034 (2018).
8. Chohan, U.: The Decentralized Autonomous Organization and Governance Issues. Notes on the 21st Century: Critical Blockchain Research Initiative, (2017).
9. Atzori, M.: Blockchain technology and decentralized governance: Is the state still necessary? Journal of Governance and Regulation (6), (2017).
10. Reijers, W. Brolcháin, F., Haynes, P.: Governance in Blockchain Technologies and Social Contract Theories. Ledger (1), 134–151 (2016).
11. Barrera, C., Hurder, S.: Blockchain Upgrade as a Coordination Game. Prysm Group (2018).
12. Arrunada, B., Garicano, L.: Blockchain: The Birth of Decentralized Governance. Pompeu Fabra University, Economics and Business Working Paper Series (1608), (2018).
13. Turnbull, S.: Corporate Governance: Theories, Challenges and Paradigms. Volume 1. SAGE, London (2008).
14. Gill, A.: Corporate Governance as Social Responsibility: A Research Agenda. Berkeley Journal of International Law 2(26), (2008).
15. Griffiths, A. B., Zammuto, R. F.: Institutional Governance Systems and Variations in National Competitive Advantage. The Academy of Management Review (30), 823–842, (2005).
16. Satidularn, C., Wilkin, C., Tanner, K., Linger, H..: Investigation of the Relationship between IT Governance and Corporate Governance. Management, Leadership and Governance, 420–423, (2013).

17. Weill, P., Ross, J. W.: IT Governance: How Top Performers Manage IT Decision Rights for Superior Results. Boston: Harvard School Press (2004).
18. Simonsson, M., Johnson, P.: Defining IT Governance – A Consolidation of Literature. EARP Working Paper MS103, Department of Industrial Automation and Control Systems, KTH.
19. Ba, S., Stallaert, J., Whinston, A. B.: Introducing a Third Dimension in Information Systems Design: The Case for Incentive Alignment. Information Systems Research (12), 225-239, (2001).
20. Holt, C. A., Roth, A. E.: The Nash Equilibrium: A perspective. National Academy of Sciences of the United States of America, 3999–4002, (2004).
21. Elsner, W., Heinrich, T., Schwardt, H., Grabner, C.: Special Issue: Aspects of Game Theory and Institutional Economics. Games (5), 188-190, (2014).
22. districtOx Education Portal, Off-Chain Governance, https://education.district0x.io/general-topics/what-is-governance/off-chain-governance/, last accessed 2020/1/4.
23. Perez, Y. B.: The Controversies of Blockchain Governance and Rough Consensus, https://thenextweb.com/hardfork/2019/01/25/the-controversies-of-blockchain-governance-and-rough-consensus/, last accessed 2020/1/4.
24. Lasaulce, S., Tembine, H.: Playing with Equilibria in Wireless Non-Cooperative Games. Game Theory and Learning for Wireless Networks (2011).
25. Nash, J.: Non-Cooperative Games. The Annals of Mathematics (54), 286-295, (1951).
26. Braggion, E., Gatti, N., Lucchetti, R., Sandholm, T.: Strong Nash equilibria and mixed strategies. arXiv (2015).
27. McKelvey, R. D., McLennan, A. M., Turocy, T. L.: Gambit: Software Tools for Game Theory, http://www.gambit-project.org, last accessed 2019/11/28.
28. Quentson, A., Miners Vote Overwhelmingly in Support of Ethereum's Hardfork, https://www.ccn.com/miners-vote-overwhelmingly-support-ethereums-hardfork/, last accessed 2019/12/18.
29. CoinMarketCap, Ethereum (ETH), https://coinmarketcap.com/currencies/ethereum/, last accessed 2019/12/15.
30. CoinMarketCap, Ethereum Classic (ETC), https://coinmarketcap.com/currencies/ethereum-classic/ratings/, last accessed 2019/12/15.